\documentclass[sn-mathphys,Numbered]{sn-jnl}
\usepackage{graphicx}%
\usepackage{multirow}%
\usepackage{amsmath,amssymb,amsfonts}%
\usepackage{amsthm}%
\usepackage{mathrsfs}%
\usepackage[title]{appendix}%
\usepackage{xcolor}%
\usepackage{textcomp}%
\usepackage{manyfoot}%
\usepackage{booktabs}%
\usepackage{algorithm}%
\usepackage{algorithmicx}%
\usepackage{algpseudocode}%
\usepackage{listings}%
\usepackage{bbold}
\usepackage{amsmath}
\usepackage{amsthm}
\usepackage{amssymb}
\usepackage{amsfonts}
\usepackage{tensor}
\usepackage{braket}
\usepackage{slashed}

\theoremstyle{definition}

\newcommand{\hl}{\mathcal{H}}

\begin{document}

\title[Relational Quantum Mechanics and Contextuality]{Relational Quantum Mechanics and Contextuality\footnote{Published in Found. Phys, Volume 54, article number 54, (2024)}}


\author*[1]{\fnm{Calum} \sur{Robson}}\email{c.j.Robson@lse.ac.uk}

\affil*[1]{\orgdiv{ Work done whilst visiting CPNSS}, \orgname{London School of Economics}, \orgaddress{\street{Houghton Street}, \city{London}, \postcode{WC2A 2AE}, \country{UK}}}

\abstract{This paper discusses the question of Stable Facts in  Relational Quantum Mechanics. I examine how the approach to quantum logic in the consistent histories formalism can be used to clarify what information about a system can be shared between different observers. I suggest that the mathematical framework for Consistent Histories can and should be incorporated into RQM, whilst being clear on the interpretational differences between the two approaches. Finally I briefly discuss two related issues: the similarities and differences between special relativity and RQM and the recent Cross- Perspectival Links modification to RQM.}

\keywords{Quantum Foundations, Relational Quantum Mechanics, Consistent Histories, Contextuality}

\maketitle

\section{Introduction}
This paper is a discussion of  Stable and Relative Facts in  Relational Quantum Mechanics. This interpretation of Quantum Mechanics was first introduced by Carlo Rovelli in \cite{Rovelli1996}. Since then, it has been challenged and developed in various ways, with growing interest in the past few years-- for example \cite{VanFraassen2010}\cite{Laudisa2019}\cite{DiBiagio2021}\cite{DiBiagio2022}\cite{Brukner2021}\cite{Pienaar2021a}\cite{Pienaar2021b}\cite{Lawrence2023}\cite{Cavalcanti2023}\cite{Mucino2022}\cite{Smerlak2007}\cite{Dussaud2019}. \\ 
The central principle of RQM is that facts about quantum systems are not observer-independent, but relative to different observers. As Rovelli says in \cite{Rovelli2022}, `Facts are \textsl{relative} to the systems that interact. That is, they are \textsl{labelled} by the interacting systems. This is the core idea of RQM'. This naturally suggests a question -- when are facts relative to one observer true relative to another?  Rovelli contrasts, `relative facts' which are only true for individual observers with, `stable facts' which are true for multiple observers\footnote{This distinction has come in for criticism, most recently in \cite{Lawrence2023}. See also the response by Rovelli et. al. in \cite{Cavalcanti2023}. }\cite{Rovelli2022} . In this paper,  I will look at how the approach to quantum logic developed by Griffiths \cite{Griffiths2008}\cite{Griffiths2014} can be used to clarify exactly which facts are stable and which are relative for which observers. I will present this, `quantum reasoning' approach to quantum logic, and explain how it is applied to families of histories of quantum states in Consistent Histories. I will then show how this mathematical framework can be given a very different physical meaning when applied to the RQM interpretation, being very clear how this differs from their use in Consistent Histories. \\
Having done this, I will comment on two other issues in the literature relating to stable facts. First, I clarify some issues about the analogy between special relativity and RQM, raised by Pienarr \cite{Pienaar2021b}. Secondly, I suggest an alternative approach to the recent, `Cross-perspectival links' modification of RQM \cite{Adlam2022}, which provides an ontological account of stable facts. \\
\section{Relational Quantum Mechanics}
In this section I will present the main points of the RQM representation, based on the recent accounts \cite{Rovelli2022} and \cite{Adlam2022}. The clarifications in Rovelli and Di Biagio's reply \cite{DiBiagio2022} to Pienaar \cite{Pienaar2021b} and Brucker \cite{Brukner2021} are also useful.  There are three general points to start. 
\begin{itemize}
	\item  There  is no ontological difference between observed and observing systems\footnote{This is in contrast to strong notions of complementarity in which there are two types of reality, classical and quantum, with properties in the  quantum reality corresponding to properties in the classical reality,  whilst not being identical to them in any straightforward way.}.  An important corollary of this is that the line between systems can be drawn anywhere-- there are many ways to divide the world into systems and subsystems, and any of these divisions is equally well described by quantum mechanics. \\
\item Though the language of, `observer' and `observing' is used, RQM does not require that observers are in any sense conscious. Any interaction between systems is a measurement within RQM. \\
\item A system cannot measure itself-- therefore any interaction must describe changes within one system from the point of view of a second system.
Rovelli \cite{DiBiagio2022} has claimed this is a consequence of certain no-go theorems (e.g. \cite{Frauchiger2018}), however, the applicability of these theorems to the issue of self-measurement has been challenged \cite{Mucino2022} \cite{Lawrence2023} and so for this paper I shall take the following statement: 
\begin{quote}
	\textbf{A1:} A system cannot measure itself, and therefore cannot ascribe a quantum state to itself 
\end{quote}
to be an axiom of RQM. 
\end{itemize}
\subsection{Outline of the Interpretation}
RQM holds that Quantum Mechanics is a theory about, `events', or, `facts', which are interactions between systems in which information is transferred\footnote{Thus the nature of the quantum world is the same as the problem with Politics according to the  mid 20th century British Prime Minister Harold Macmillan-- `Events, Dear Boy-- Events!'} . Quantum Mechanics is then a tool for predicting probabilities of future events based on past events. As Rovelli says in \cite{Rovelli2022}:
\begin{quote}
	RQM interprets QM as a theory about physical \textsl{events}  or \textsl{facts}. The theory provides transition amplitudes of the form $W(b,a)$ that determine the probability $P(b,a)=\lvert W(b,a)\rvert^{2}$ for a fact (or collection of facts) $b$ to occur, given that a fact (or collection of facts) $a$ has occured... The insight of RQM is that the transition amplitures $W(b,a)$ must be interpreted as determining physical facts only if the physical facts $a$ and $b$ are relative to the same system. 
\end{quote}
 In recent presentations \cite{DiBiagio2021}\cite{Adlam2022}, the following set of postulates is used. I have copied this verbatim from \cite{Adlam2022}:
\begin{quote} 
	\textbf{R1: Relative Facts.} Events, or Facts, can happen relative to any physical system\footnote{This idea of a, `Relative State' is taken by Rovelli from the Everettian interpretation of Quantum Mechanics, though given a different physical meaning}  \\ \\ 
	\textbf{R2: No Hidden Variables.} Unitary quantum mechanics is complete\\ \\ 
	\textbf{R3: Relations are Intrinsic. } The relation between any two systems $A$ and $B$ is independent of anything that happens outside these systems' perspectives\\ \\ 
	 \textbf{R4: Relativity of Comparisons.} It is meaningless to compare the accounts relative to any two systems except by invoking a third system relative to which the comparison is made\\ \\ 
	 \textbf{R5: Measurement.} An interaction between two systems results in a correlation between these systems and a third one: that is, with respect to a third system $W$, the interaction between two systems $S$ and $F$ is described by a unitary evolution that potentially entangles the quantum states of $S$ and $F$. \\ \\
	 \textbf{R6: Internally Consistent Descriptions.} In a scenario where $F$ measures $S$, and $W$ also measures $S$ in the same basis, and $W$ then interacts with $S$ to, `check the reading' of a pointer variable (i.e. by measuring $F$ in the appropriate, `pointer basis'), the two values are found in agreement. 
\end{quote}
All this means that RQM is highly contextual. This is the opposite of being noncontextual, which is defined by the Stanford Encyclopedia of Philosophy \cite{SEPKS}  as the condition that
\begin{quote}
	If a quantum mechanical system possess a property (value of an observable) then it does so independently of any measurement context (i.e. independently of \textsl{how} that value is eventually measured)
\end{quote}
Conversely, contextuality occurs when the properties a quantum system possesses are dependent on the context of the measurement\footnote{We can find examples of this without needing to invoke any quantum effects.  Suppose we measure the temperature of a pan of water. The thermometer gets a reading by taking heat from the pan, and therefore the temperature measured by the thermometer is not the exact temperature of the pan before the measurement, but will be slightly less. For usual applications, this does not matter-- the difference in heat will be negligable, probably far less than the precision of the thermometer. But now suppose the thermometer is large relative to the size of the pan. Then the thermometer will draw so much heat out of the pan in the measurement process that the reading on the thermometer is  not an accurate measurement of what the temperature was before the measurement took place. \\
This is where Quantum Theory comes in. We can try and fix this by using a smaller thermometer, but Quantum Theory seems to indicate that there is a smallest length scale to everything (and a smallest energy, temperature, etc.). Therefore there will come a point where we cannot decrease the size of the measuring instrument any further relative to what we are measuring. This means that our measurement will no longer reveal the properties of the system before the act of measurement. Instead we can only measure what the system has become after we have interacted with it. This point is strongly made by Dirac in the introduction to \cite{Dirac1988}.\\
One of my students suggested an alternative and rather lovely example of Contextuality. If you're feeling down and someone asks how you are, the answer to the question is usually going to be happier than you were before it was asked. }.  RQM states that, `facts', including measurement probabilities, are relative to potential observers. This is inherently contextual since the, `facts' pertaining to a system depend upon which observer is measuring, and hence upon a particular context. 
\subsection{The 1996 Axioms  and Cross-Perspectival Links}
As well as this standard set of principles, there are some alternative descriptions of RQM which are relevant for this paper. First, Rovelli's original paper \cite{Rovelli1996} introduces a pair of axioms explaining RQM in terms of information:
\begin{quote}
	\begin{enumerate}
		\item There is a maximum amount of relevant information which can be extracted from a system
		\item It is always possible to acquire new information about a system. 
	\end{enumerate}
\end{quote}   
As Adlam and Rovelli say \cite{Adlam2022}, `together they imply that sometimes when an agent acquires new information about a system, some of their previous information becomes irrelevant'
\footnote{ For example, there will be some measurements (eg $x$ and $p_{y}$) which can be mutually stable because the act of measurement one does not affect the system in ways which disturb the value of the other properties.\\
	However, other measurements (say, $x$ and $p_{x}$) are such that measuring one does disturb the value of the other. In order to measure the position of a particle, I must contact it, e.g. by hitting it with a photon. This will affect its momentum. If I try and measure its momentum first, then this requires interaction with some medium over a period of time, which will affect the final position relative to where it would have been had I not measured the momentum first. So the position after the momentum measurement cannot be taken as the position at that time had the momentum measurement not taken place, and vice versa. This is the contextuality property in action. See also Feynman's analysis of the double slit experiment in \cite{FeynmannLect}. The relation between different measurements and their properties is give by the relation between operators on the Hilbert space, in another example of Wigner's, `unreasonable effectiveness of mathematics' \cite{Wigner1960}}.
As is pointed out in \cite{Adlam2022}, this presentation has not been used much in the subsequent development of RQM, but it will be conceptually useful in this paper-- the mathematical framework of Consistent Histories, as I will show, can be a way to determine exactly which measurements destroy previous information about the system, and which preserve it. \\
Second, in a recent paper, Adlam and Rovelli \cite{Adlam2022} have introduced a new axiom for RQM, which they call \textsl{Cross-perspectival links (CPL)}. 
\begin{quote}
	In a scenario where some observer Alice measures a variable $V$ of a system $S$, then provided Alice does not undergo any interactions which destroy the information about $V$ stored in Alice's physical variables, if Bob subsequently measures the physical variable representing Alice's information about the variable $V$, then Bob's measurement result will match Alice's measurement result. 
\end{quote}
This is in response to the challenge that the original formulation of RQM implies that each observer has their own solipsistic description of the universe. By \textbf{R4}, we need to introduce a third observer, Charlie to check if Alice and Bob have measured the same value of variable \textsl{V}.  But then whilst Charlie will always measure Alice and Bob as having a consistent description of an event S, via \textbf{R6 }, he apparently has no way of knowing if this description was the one which either Alice or Bob had before the measurement. This implies each observer lives in a solipsistic universe and has no way to check their findings with any other observer.  The new axiom is meant to counter this objection by guaranteeing that different observers will agree on measurement results provided that the information held by each observer is not destroyed when they check each other's results. I broadly agree with this approach, and will comment in more detail later  in the paper. \\
For now, one advantage of this axiom is that it allows us to be precise about what is meant by a system in RQM. Since RQM is a theory about events and facts, it may not be obvious what the ontological status of Systems is. The CPL approach allows us to view a system as emergent out of some stable pattern of events. To quote from \cite{Adlam2022}, `A System can simply be identified with a set of quantum events which are related to one another in certain lawlike ways, as captured by the formalism of Quantum Mechanics'. 
\subsection{The Wavefunction}\label{sec:wavefunction}
Facts about a system are contained in the wavefunction describing a system. We therefore next turn to the question of how the  wavefunction is understood in RQM. 
The wavefunction an observer $\mathcal{O}$ ascribes to a system $ \mathcal{S}$ is a statement of the facts that $\mathcal{O}$ ascribes to $\mathcal{S}$. In general, we have:
\begin{itemize}
	\item If a system $\mathcal{S}$ is in an eigenstate $\ket{\phi_{i}}$ relative to a system $\mathcal{O}$, then with probability 1 $\mathcal{O}$ will measure $\mathcal{S}$ to be in that eigenstate. Unless otherwise indicated we can therefore view $\mathcal{S}$ as physically being that eigenstate. 
	\item If a system $\mathcal{S}$ is a superposition $\sum_{i} a_{i}\ket{\phi_{i}}$ relative to a system $\mathcal{O}$, then this means $\mathcal{O}$ will measure the eigenstate $\ket{\phi_{i}}$ with probability $\lvert a_{i}\rvert^{2}$ . No other conclusions can be drawn about the physical state of $\mathcal{S}$\footnote{For a discussion of some of the mathematical issues involved, see \cite{Oldofredi2021}}. 
\end{itemize}
This implies the key feature of RQM-- that different observers can assign different wavefunctions to the same system\footnote{We could consider the possibility that the properties assigned to the observed system are subjective, and a system has different properties relative to different observers due to the different observers assigning different properties for subjective reasons (for example, they may only have experimental knowledge of certain properties, or may have different degrees of confidence in the reliability of their equipment). This, however, is precisely the QBism interpretation of Quantum Mechanics. This is based on a subjectivist interpretation of probability theory, and assumes that the probabilities ascribed to a state are subjective and depend on an individual's own beliefs about the probability, rather than being, for example, linked to the objective frequency of the measurement outcomes. The similarities and differences between RQM and QBism are expertly discussed in \cite{Pienaar2021a}, and the difference between them can be summed up by the fact that for RQM, an observed system being in a definite eigenstate (probability 1) relative to an observer is a statement about the observed system, that it has a definite physical property, whereas in QBism, it is a statement about the beliefs of the observer, that they will certainly measure the system to be in that eigenstate. As Pienaar \cite{Pienaar2021a}  points out, this has the important consequence that whereas in RQM, an observing system  does not have to be a conscious being, in QBism by definition an observer must be conscious (i.e. able to have beliefs about probabilities). }. 
Consider the following example. We have a source of particles which we know produces particles with random polarization.  Alice places a vertical polarizer and detector in front of this source, and measure the spin from each particle after passing through the polarizer. She will measure either spin up or spin down from each particle. \\
After being produced, and before the measurement, she can assign a particle the wavefunction $\frac{1}{\sqrt{2}}\ket{\uparrow}+ \frac{1}{\sqrt{2}}\ket{\downarrow}$ . Viewing the wavefunction as an epistemic object, this means that she knows that the particle will be detected either as $\ket{\uparrow}$ or $\ket{\downarrow}$, each with probability $1/2$ \footnote{This process is inherently stochastic. Rovelli suggests  that some of the indeterminacy comes from the fact that observers can never fully describe their own role in the measurement (since in RQM we cannot assign a quantum state to ourselves), and part is inherent to the nature of quantum reality.  An interesting and relevant discussion of the distinction in quantum theory between intrinsically quantum probabilities and probabilities based on epistemic ignorance is given in \cite{Penrose1971}}.  There is no other physical meaning to this superposition \footnote{Unlike, e.g. physical collapse interpretations of the wavefunction}. \\
What about after the measurement?  Because we are passing it through a vertical polarizer, then there is a definite (and stochastic) change to the particle, which leads to it having one of the two eigenstates $\ket{\uparrow} $ or $\ket{\downarrow}$. This eigenstate now has the epistemic meaning that we know the particle is in that eigenstate. In this case, we can conclude that the particle is physically rotating with the measured spin. So the epistemic change also corresponds-- in this particular case-- to a physical change. \\
Now Bob wants to measure particles from our source with a horizontal polarizer. He would assign the particle the state $\frac{1}{\sqrt{2}}\ket{\leftarrow}+ \frac{1}{\sqrt{2}}\ket{\rightarrow}$, since he expects to find either $\ket{\leftarrow}$ or $\ket{\rightarrow}$ with probability 1/2. The fact that Alice assigns the wavefunction $\frac{1}{\sqrt{2}}\ket{\uparrow}+ \frac{1}{\sqrt{2}}\ket{\downarrow}$ and Bob assigns the wavefunction $\frac{1}{\sqrt{2}}\ket{\leftarrow}+ \frac{1}{\sqrt{2}}\ket{\rightarrow}$ is not a contradiction in RQM, because the wavefunction represents the information each has about the particle, and there is no reason for this to be identical. \\
Even if Alice has performed her measurement, and knows the particle is in state $\ket{\uparrow}$, it is not a contradiction for Bob to assign the particle the state $\frac{1}{\sqrt{2}}\ket{\leftarrow}+ \frac{1}{\sqrt{2}}\ket{\rightarrow}$, since it is still true that he will measure either  $\ket{\leftarrow}$ or $\ket{\rightarrow}$ with probability 1/2. 
\subsection{Stable and Relative Facts}
As stated earlier, once we have the picture of multiple observers with their own descriptions (sets of facts) of a system, we want to  ask whether any elements of these descriptions are shared. Rovelli and Di Biagio give an analysis of this in \cite{DiBiagio2022}, distinguishing between stable facts and relative facts. Given a set $\mathcal{O}$ of observing systems, and an observed system $\mathcal{S}$, stable facts are those which are assigned by all the systems in $\mathcal{O}$ to $\mathcal{S}$, whereas relative facts are only assigned to $\mathcal{O}$ by some, or even just one, of the systems in $ \mathcal{O}$. \\
It follows that stable facts obey the standard conditional probability law
\begin{equation}\label{eqn:stable}
	P(a_{j})=\sum_{i\neq j}P(a_{j}|a_{i})P(a_{i})
\end{equation}
which is taken by Di Biagio and Rovelli to define stable facts. They go on to define relative facts:
\begin{quotation}
	Relative facts are defined to happen whenever one physical system interacts with another system
\end{quotation}
So whenever an observing system $\mathcal{O}_{1}$ interacts with an observed system $\mathcal{S}$, and gains information about it, that information is a relative fact between $\mathcal{O}_{1}$ and $\mathcal{S}$.   Suppose there is another  observer $\mathcal{O}_{2}$, which also has relative facts about $\mathcal{S}$. Let $\{a_{i}\}$ be a set of relative facts where each fact $a_{i}$ can be relative to either $\mathcal{O}_{1}$ or $\mathcal{O}_{2}$. If each  $a_{i}$ satisfies equation (\ref{eqn:stable}) then each fact in the set $\{a_{i}\}$ is stable for $\mathcal{O}_{1}$ and $\mathcal{O}_{2}$-- that is, it is a true fact for both systems.  \\
 RQM appeals to decoherence to explain why the world we see (on a classical scale) seems to be made up of stable facts which are observer-independent. Rovelli \cite{Rovelli2022} also points out that most other interpretations of Quantum Mechanics implicitly only deal with facts which are stable, because they do not use the RQM principle that facts are relative to particular observers.\\
This definition, using equation (\ref{eqn:stable}), is hard to check for specific examples, and so stable facts remain an area of debate in the analysis of RQM. However, I think we can give a more rigorous account of stable facts using the the logical framework developed by Griffiths \cite{GriffithsSEP} known as \textsl{Quantum Reasoning}, as part of the Consistent histories interpretation.  To do this, I will need to discuss the consistent histories formalism of quantum mechanics, which will provide the necessary mathematical framework.  This will be the main argument of this paper.
\section{RQM and Consistent Histories}\label{sec:CH}
The Consistent Histories (CH) formulation of Quantum Mechanics is based on the work of Griffiths \cite{Griffiths1984}, Omnes \cite{Omnes1987}, and Gell-Mann and Hartle\cite{GellMan1990}. Each of these figures has contributed to the mathematics of the interpretation, however they have differing perspective on the physical conclusions to be drawn-- for a history and discussion, see \cite{Rocha2022}.  The description of the CH interpretation given here mainly follows Griffiths, being a summary of \cite{GriffithsSEP}, supplemented by \cite{Griffiths2008}-- though I should clarify that the physical meaning I am attaching to the histories is very different from the standard one, as I shall discuss below. \\
In this section, closely following \cite{GriffithsSEP}, I will approach the CH formalism as defining a non-standard logic on quantum sample spaces, which Griffiths terms \textsl{Quantum Reasoning}. This is a consequence of the non-commutativity of quantum operators. Next, I will discuss how this is extended to histories, or chains of operators at different times. \\ 
After this overview, I will suggest how this quantum logic can be used to give a method for determining whether facts shared by different observers are stable or not (in the RQM sense). Finally, in a concluding section, I make clear how the use of Quantum Reasoning I am suggesting for RQM differs from its original intended use within the Consistent Histories formalism. 

\subsection{The Consistent Histories Formulation}
There are two major features of the Consistent Histories approach, following Griffiths. The first is that Quantum Mechanics is fundamentally stochastic. The second is that the noncommutative nature of the operators in Quantum Theory necessitates a special kind of logical framework known as, `quantum reasoning'\cite{Griffiths1996} \cite{Griffiths2014}. This is implemented by treating the Hilbert Space as the quantum analogue of a classical Phase Space as I shall now outline. \\
In Classical Mechanics, the possible values of our variables can be labelled as points in a phase space. Suppose we are considering a set of values corresponding to a region $\gamma$ of the phase space. We can define indicator functions which pick out that set of points by 
\begin{equation}
	P(x)= 
	\begin{cases}
		1 &  \ x\in\gamma
		\\ 0  & \ x\in\gamma^{c}
	\end{cases}
\end{equation}
By this construction the phase space splits neatly into $\gamma$ and $\gamma^{c}$, where $\gamma^{c}$ is the complement of $\gamma$. \\
In the Quantum case, our quantum phase space is a Hilbert Space $\mathcal{H}$. Properties now correspond to subspaces of $\hl$, which are picked out by projection operators $\mathcal{P}$.
\begin{equation}
	\mathcal{P}_{\Gamma}\ket{\phi}=
	\begin{cases}
		1 & \ket{\phi}\in \Gamma \\
		0 & \ket{\phi}\notin \Gamma 
	\end{cases}
\end{equation}
We define the projector $\neg\mathcal{P}$ to be $\mathbb{1}-\mathcal{P}$. Now we come to the key difference between the classical and quantum cases. In the classical case, every point lies either in $\Gamma$ or $\Gamma^{c}$. In the quantum case this no longer applies. Suppose  $\hl=\text{Span}\big(\ket{\phi_{1}}, \ \ket{\phi_{2}}\big)$ and $\Gamma$ is the subspace $\text{Span}\big(\ket{\phi_{1}}\big)\subset\hl$. Then the state vector $\ket{\phi_{1}}+\ket{\phi_{2}}$ is neither in $\Gamma$ or $\Gamma^{c}$. How can we interpret such states?\\ \\ 
This is an example of a larger issue. Suppose we have two subspaces, $P$ and $Q$. In classical terms, $P\wedge Q$ is an intersection of $P$ and $Q$, but this is ambiguous in Quantum theory. This is because the projectors corresponding to $P$ and $Q$ might not commute, and so it is ambiguous whether we should choose $\mathcal{P}\mathcal{Q}$ or $\mathcal{Q}\mathcal{P}$ as the appropriate projector for $\mathcal{P}\wedge\mathcal{Q}$. \\
Consistent Histories attempts to solve this by defining the projector $\mathcal{P}\mathcal{Q}$ iff $\mathcal{P}$ and $\mathcal{Q}$ commute. Otherwise it says that $P\wedge Q$ is meaningless
\footnote{Another strategy was adopted by Von Neumann in his 1930s development of, `Quantum Logic' \cite{Neumann1936}}.
This gives a logic which no longer follows the principle of the excluded middle-- it allows a proposition to be three things; either true, false or non- defined. That said, as we shall see in the next section we can only draw conclusions when the sample space contains no undefined propositions. 
\subsection{Probabilities and Quantum Reasoning}
Once we have the Quantum phase space, along with the consistent histories logic, we can use it to calculate probabilities. We can think of a probability theory as giving a triple $\big( \mathcal{S}, \mathcal{E}, \mathcal{M}\big)$ where
\begin{itemize}
	\item $\mathcal{S}$ is a sample space- the underlying objects or situations we are working with 
	\item $\mathcal{E}$ is an event algebra-- the set of actual events involving elements of the sample space we want to know the probability of. The algebra structure is given by unions and intersections. 
	\item $\mathcal{M}$ is a probability measure
\end{itemize}
For classical physics, the sample space is given by regions of the phase space, and the event algebra is given by the corresponding indicator functions. In the quantum case, each division of the Hilbert Space into subspaces gives a different sample space, and the event algebra is given by the projectors onto those subspaces. As these examples show, there is a one-to-one relation in these cases between the event algebra and sample spaces, and so we shall refer to them interchangeably. In both the quantum and classical cases, the probabilities are usually not given intrinsically and are assigned via theoretical or empirical considerations. \\
The main different between the quantum and classical cases, according to the consistent histories interpretation, is that in the classical case, the event algebra is well defined for any set of subspaces, so in practice we can compare events in different probability frameworks, or change the probability framework that we are using, without any problems. \\
in Quantum Mechanics on the other hand\footnote{at least according to consistent histories}, the event algebra is only well defined when the projectors commute (otherwise we define expressions like $P\wedge Q$ to have no meaning, as explained above). Therefore we cannot straightforwardly add new events into the event algebra, or compare probabilities between different frameworks. We must check to make sure that the events and frameworks which we are comparing give a well defined event algebra. \\
We say that two sample spaces (projective decompositions) of a Quantum operator are compatible if all the projectors of one sample space commute with the projectors in the other. If $\{\mathcal{P}_{i}\}$ and $\{\mathcal{Q}_{i}\}$ are the event algebras, then we require
\begin{equation}\label{eqn:con1}
	\mathcal{P}_{j}\mathcal{Q}_{k}=\mathcal{Q}_{k}\mathcal{P}_{j} \ \ \forall j, k
\end{equation}
If the frameworks are compatible, then we can combine them\footnote{Sometimes this involves a process called refinement, which I am not discussing in this brief introduction-- see \cite{Griffiths2008} or \cite{GriffithsSEP}}. Otherwise, if the frameworks are incompatible, then we adopt the \textsl{Single Framework Rule}, which is the central feature of the Consistent Histories interpretation. \\
This states that we cannot directly compare probabilities from incompatible frameworks-- we must only use probabilities from a single compatible framework \footnote{Though this framework may be made of of several compatible frameworks joined together-- see below}. Finally, we can define measurement operators in the usual way as combinations of projectors. These operators then form the event algebra. For some worked examples of compatible and incompatible frameworks, see section 3.4 of \cite{GriffithsSEP}.
\subsection{Histories}
So why is the interpretation called, `Consistent Histories'? That comes about because we want to compare measurement at different times, and to discuss sequences of measurements. We first define a time-graded Hilbert space for successive times $t_{0}, t_{1}..., t_{n}$ as
\begin{equation}
	\mathcal{H}=\mathcal{H}_{0}\odot\mathcal{H}_{1}\odot...\odot\mathcal{H}_{n}
\end{equation}
where the $\odot$ represents a tensor product-- the different symbol being used as a reminder that each Hilbert space and its corresponding projectors is associated with a different time. \\
We then define a Quantum history as a tensor product of projectors for each fixed-time Hilbert Space:
\begin{equation}
	Y^{a}=P_{0}\odot P_{1} \odot... \odot P_{n}
\end{equation}
These $Y^{a}$ form a sample space so long as $\sum P_{i} =\mathbb{1}_{i}$ for each $\mathcal{H}_{i}$. In this case, $\sum Y^{a}=\tilde{\mathbb{1}}$, where $\tilde{\mathbb{1}}$ is identity on $\mathcal{H}$-- the tensor product of the identities $\mathbb{1}_{i}$ on the fixed time Hilbert Spaces. This defines our Event Algebra and Sample Space\footnote{Remember these are dual since defining a sample space of subspaces automatically picks out an event algebra based on their projectors, and vice versa}.
The remaining task is to assign probabilities to the histories $Y^{a}$. To do this, we construct \textsl{Chain kets}
\begin{equation}
	\ket{Y^{a}}=P_{n}T(t_{n}, t_{n-1})P_{n-1}T_(t_{n-1}, t_{n-2})... T(t_{2}, t_{1})P_{1}T(t_{1}, t_{0})P_{0}
\end{equation}
Note that this is an operator on a single Hilbert space at time $t_{n}$. The operators $T_(t_{i}, t_{i-1})$ are unitary operators which describe time evolutions. What they is can vary according to the system described. Usually, they will be given by the Schroedinger equation. If we do not wish to take into account time evolution of a state then we can just set $T(t_{i}, t_{i-1})$ to be the identity operator. We then assign each history the probability
\begin{equation}
	Pr(Y^{a})=\braket{Y^{a} | Y^{a}}
\end{equation}
provided that all the histories projectors satisfy the consistency condition
\begin{equation}\label{eqn:con2}
	\braket{Y^{a} | Y^{a'}}=0\ \text{for} \ a \neq a'
\end{equation}
We call such a set $\{ Y^{a} \}$ a consistent family of histories.  It defines a compatible framework at time $t_{n}$. We can think of $\ket{Y^{a}}$ as a series of rotations/ dilations from the unitary operators $T(t_{i}, t_{i-1})$ and projections from the $P_{i}$, which together make up a single operator which turns the initial subspace corresponding to $P_{0}$ to a final state corresponding to the projector $Y^{a}$. \\
Note that states with probability zero now include those where the dynamics are impossible-- i.e. where the combinations of rotations, dilations and projections lead to the empty set. \\ \\
As a final aside, we can use this to understand the Heisenberg uncertainty principle. The operator whose eigenstate is a position measurement $x_{0}$ is is the delta function $\delta (x-x_{0})$, and the operator corresponding to a measurement of the momentum of an object is $( ih)d/dx$, evaluated at the point $x_{0}$. These operators are not compatible at the same point $x_{0}$, and so there is no meaning to assigning exact values of both position and momentum to a particle at point $x_{0}$ \footnote{From the point of view of Quantum Reasoning, this is presumably the solution to Zeno's paradoxes. The construction of the paradoxes involves inconsistent operators, and therefore incompatible frameworks, in the sense of the next section. It would be interesting and instructive to show this rigorously for each paradox}. 
\subsection{Stable facts and compatible frameworks}
Now we have given the mathematical details of the CH formalism, we can apply it to RQM. This will make it clearer what RQM means by, `facts' and in particular it can be used to give a more rigorous definition of, `stable facts'. \\
I suggest that a history is taken to be the set of facts held by an observer about the system it describes, at different time steps. We can then assign probabilities to these histories. This demonstrates Rovelli's principle that \cite{Rovelli2022}
\begin{quote}
	[Quantum] theory determine(s) the probability... for a fact (or collection of facts) $b$ to occur given that a fact (or collection of facts) $a$ has occurred.
\end{quote}
 Consistent Histories also gives a precise way of talking about stable and relative facts between systems. Given two consistent families, CH gives a criterion for determining when the two families can be combined into a single framework, and an algorithm for how to do so. If we associate the two families with the facts determined by each system, then we can say that the facts are stable relative to both systems if the families associated with those facts are compatible. \\
Stable facts are those which come from a consistent family of histories. Within a given family of histories, the event algebra automatically follows the rules of classical probability. So Rovelli and Di Biagio's \cite{DiBiagio2022} definition of a relative fact as one for which
\begin{equation}
	P(b)=\sum_{i}P(b|a_{i})P(a_{i})
\end{equation}
holds is always satisfied. This gives a well defined condition for which facts are stable and which are not. \\
So what is this condition? Given two consistent families of histories, $\{K_{i}\}$ and $\{Y_{j}\}$, we can combine them into a single family of histories iff the following two conditions are satisfied:
\begin{enumerate}
	\item The operators for each family at each time step $t_{i}$ must commute
	\item The family of histories $\{K_{i}Y_{j}\}$, given by taking the tensor product of $\{K_{i}\}$ and $\{Y_{j}\}$ at each time step, must be a consistent family.  Essentially these families represent $\{ K_{i}\wedge Y_{j}\}$, which are well-defined due to condition 1
\end{enumerate} 
Therefore if the $\{K_{i}\}$ correspond to the interactions with (and hence relative facts about) $S$ by two observers $\mathcal{O}_{i}$, the facts are stable for the $\mathcal{O}_{i}$ if the corresponding families of histories are compatible.  These conditions make physical sense in the RQM context-- condition 1 means that the measurements made by each subsystem at each time do not interfere with one another, whereas condition 2 guarantees that the pairs of histories considered together define a consistent space of probabilities. I will now give some examples to illustrate this. Note that in order to check compatibility, we must consider the whole sample space, and must therefore include all the histories which could have been measured given the experimental setup, not just the one result which was actually obtained.  \\
First, suppose that we have two observing systems $\mathcal{O}_{1}$ and $\mathcal{O}_{2}$ measuring the same system $\mathcal{S}$. They each attribute the same initial state, $\ket{\psi_{0}}$, to $\mathcal{S}$ at time $t_{0}$ and they both measure the spin of $\mathcal{S}$ in the $x$ direction at time $ t_{1}$. At time $t_{2}$, $\mathcal{O}_{1}$ measures the momentum eigenstate in the $x$ direction of $\mathcal{S}$, and $O_{2}$ measures the momentum eigenstate in the $y$ direction. Assume all time evolution is unitary, following Schroedinger's equation. Then the two families of histories are\footnote{I am making a simplification here- technically we should add in histories with operators of the form $\mathbb{1}-\sum o_{i}$ at each time step (where $o_{i}$ are the projection operators at each time step) to ensure that the sum of the projectors at each time step gives the identity}
\begin{align}\nonumber
	K_{i}&=\ket{\phi_{0}}\odot\sigma_{x,k}\odot p_{x,l}\\
	Y_{j} &=\ket{\phi_{0}}\odot \sigma_{x,m}\odot p_{y,n}
\end{align} 
Each history represents one possible pair of measurement outcomes at $t_{1}$ and $t_{2}$. Here $\sigma_{x,k}$ is eigenstate $l$ of the operator $\sigma_{x}$ and so on.  As mentioned above, I am summing over all the possible measurement outcomes, even though only one of these outcomes will be measured for both $\mathcal{O}_{1}$ and $ \mathcal{O}_{2}$ because it is necessary to check that the sample space of all possibilities is consistent. \\
The operators at each time value commute, and we can easily show that the pairs of operators  form a consistent family due to the orthogonality of the operators in each family at $t_{2}$\footnote{See chapter 11 of \cite{Griffiths2008} for a guide to techniques for showing consistency}. Therefore we can combine these operators into a single framework. This means that the facts relative to $\mathcal{O}_{1}$ and $\mathcal{O}_{2}$ are stable at each timestep, so they can agree on which properties $\mathcal{S}$ has at each $t_{i}$. \\
What would be an incompatible pair of families? Suppose that $\mathcal{O}_{1}$ still measured the spin in the $x$ direction at time $t_{1}$, but $\mathcal{O}_{2}$ now measures the spin in the $y$ direction at $t_{1}$. Then the two families of histories
\begin{align}\nonumber
	K_{i}&=\ket{\phi_{0}}\odot\sigma_{x,k}\odot p_{x,l}\\
	Y_{j} &=\ket{\phi_{0}}\odot \sigma_{y,m}\odot p_{y,n}
\end{align} 
are incompatible because the projectors $\sigma_{x}$ and $\sigma_{y}$ at time $t_{1}$ do not commute. Therefore $\mathcal{O}_{1}$ and $\mathcal{O}_{2}$ have only relative facts, and not stable ones. \\
This  use of Quantum Reasoning in RQM is a way of making precise the 1996 axioms
\begin{quote}
	\begin{enumerate}
		\item There is a maximum amount of relevant information which can be extracted from a system
		\item It is always possible to acquire new information about a system. 
	\end{enumerate}
\end{quote}   
Compatible histories do not interfere with or destroy the information obtained through their measurements, whereas incompatible ones do. The mathematical framework of Consistent Histories is therefore a powerful tool for investigating Relational Quantum Mechanics. That said, these are elementary examples, and further work is needed to explore the behaviour of stable facts, and potentially to improve this algorithm for determining them. For example, is there a way to have stable facts not only at a single time, but across multiple times? Additionally, if the facts are stable until $t_{i-1}$ is there a way to, `regain' stability of facts from a later time $t_{i+1}$ after losing it at $t_{i}$? These are topics I hope to explore in future work. A useful way to begin this would be to analyse the various quantum paradoxes using the quantum reasoning method applied to RQM, as outlined in this paper, and comparing the results to the analyses in the standard CH framework-- see, for example, the later chapters in \cite{Griffiths2008}.
\section{Differences between RQM and CH}
Griffiths has indicated in private correspondence that he does not consider the use I have made of his theory of Quantum Reasoning to be valid. I therefore want to make clear how the way I am using Quantum Reasoning differs from its intended use within the context of Consistent Histories. I shall first discuss the conceptual difference between the two interpretations, before giving some examples through discussing measurement in both theories. 
\subsection{Differences in interpretation}
The main difference between my approach in this paper, and Griffith's original approach \cite{Griffiths2008}\cite{Griffiths2014} is the physical interpretation attached to the choice of framework. For Griffiths, the choice of framework is determined by the following four principles, quoted verbatim from \cite{GriffithsSEP}:
\begin{quote}
	 \textbf{R1: Liberty}. The physicist is free to employ as many frameworks as desired when constructing descriptions of a particular quantum system, provided the principle \textbf{R3} below is strictly observed.\\
	\textbf{R2: Equality}. No framework is more fundamental than any other; in particular, there is no `true' framework, no framework that is `singled out by nature'.\\
	\textbf{R3: Incompatibility}. The single framework rule: incompatible frameworks are never to be combined into a single quantum description. The (probabilistic) reasoning process starting from assumptions (or data) and leading to conclusions must be carried out using a single framework.\\
	\textbf{R4:	Utility}. Some frameworks are more useful than others for answering particular questions about a quantum system.
\end{quote}
The idea here is that any probability framework is just as good as any other probability framework-- we just need to be sure to only use a single framework with a consistent event algebra. In particular, there is no physical meaning attached to any choice of framework-- all are equally valid. In \cite{Griffiths2014}, he says that different frameworks reveal different \textsl{aspects} of reality, and that the fact that a single quantum reality can be described by different (and possibly incompatible) frameworks is the new nature of things which quantum mechanics reveals to us.\\
Therefore the move from the classical to the quantum world involves the abandonment of the idea that there is a single correct description of the universe (Griffiths calls this idea the \textsl{Principle of Unicity}) Instead, there are multiple correct (and incompatible) descriptions of the same events. Griffiths uses the classical analogy of looking at different properties of an object in order to answer different questions about it (e.g. the capacity of a mug and its material) but is clear that this is just an analogy-- the different aspects on the quantum level are more fundamental (as can be seen from the fact that the families used to describe them can be incompatible). \\
This approach is similar to RQM's statement that there is not a unique description of the universe, but that there is a description relative to each observer. In RQM the different descriptions of the universe come about from the different observers and their different interactions with the universe. In Consistent Histories, the different descriptions are given relative to different frameworks, which are different, observer-independent, descriptions of the same system (in principle any observer could use any framework). This is what motivates my approach in this paper to assign different frameworks to different observers and their histories of interactions. The different frameworks in CH become different potential sets of interactions between observer and system in RQM. 
\subsection{Measurements in CH}
A good example of the distinction between these approaches comes from the CH treatment of measurement. As outlined in \cite{Griffiths2017}, a key feature of CH is that it denies any special status to measurement as opposed to any other kind of event. RQM also denies that measurements have a special ontological status-- every interaction between two systems is a measurement of each by the other-- but it does give measurements a special epistemic status relative to an observer, since the measurement updates the observer's description of the measured system. \\
In CH, we can model a measurement in the following way \cite{Griffiths2017}. Let $\ket{\phi_{0}}$ be the initial state of the measured system, and let $\ket{M_{0}}$ be the initial state of the detector. Suppose we can decompose $\ket{\phi_{0}}$ into eigenstates $\ket{s_{i}}$, \& let $\ket{M_{i}}$ be the state in which the detector measures $\ket{s_{i}}$. Finally, let $[Y_{i}]$ denote the projector corresponding to the state $\ket{Y_{i}}$, and $\{[Y_{i}]\}$ indicate one history for each choice of $[Y_{i}]$\\
Now, consider the following family of histories, describing a measurement, with appropriate time evolution:
\begin{equation}\label{eqn:fam1}
	[\phi_{0}]\otimes [M_{0}]\odot\big\{[s_{i}]\otimes [M_{0}]\big\}\odot\big\{[s_{i}]\otimes M_{i}\big\}
\end{equation}
This framework allows us to conclude that the conditional probability of the system being in state $s_{i}$ at time $t_{1}$ given that we measured $M_{i}$ at $t_{2}$ as
\begin{equation}
	P\big([s_{i}]_{t_{1}}|M_{j, t_{2}}\big)=\delta_{ij}
\end{equation}
If we measured $M_{i}$ at $t_{2}$ we can conclude with probability 1 that the combined state of the system and detector at $t_{1}$ was $\ket{s_{i}}\otimes M_{0}$. Now consider a different family
\begin{equation}\label{eqn:fam2}
	\big([\phi_{0}]\otimes [M_{0}\big)]\odot\big([\phi_{0}]\otimes [M_{0}]\big)\odot\big\{[s_{i}]\otimes[M_{i}]\big\}
\end{equation}
Since this family does not include the states $\ket{s}_{i}$ at $t_{1}$, we can say nothing about whether or not the system was in one of these states at that time. Here we come to the difference between the application of Quantum Reasoning to RQM which I have outlined in this paper, and its original use in CH. In CH, we are free to choose any framework to describe a system. So even if we cannot draw any conclusions about whether the system was in one of the states $s_{i}$ from the family (\ref{eqn:fam2}), we can just as well use the family (\ref{eqn:fam1}) to draw such a conclusion. \\
In RQM, on the contrary, I am suggesting that each history has a physical interpretation as referring to  the observer's history of interactions (and hence knowledge of the system being described at each time step). Therefore family (\ref{eqn:fam1}) represents a situation in which the observer knows that the system is definitely in one of the states $\ket{s_{i}}$ at $t_{1}$. In this case, measuring $\ket{M_{i}}$ at $t_{2}$ allows us to conclude that the system was in state $s_{i}$ at $t_{1}$. \\
However, family (\ref{eqn:fam2}) represents a case where the observer does not know that the system is definitely in the state $\ket{\uparrow}$ or $\ket{\downarrow}$, only that it has probability 1/2 to be found in either state under an appropriate measurement.\\
In RQM these are two different situations, and we are not free to choose between them. The first family would represent, for example,  a situation where we know our initial state $\ket{\phi_{0}}$ comes from a machine which will produce either  $\ket{\uparrow}$ or $\ket{\downarrow}$, whereas the second family could represent a situation where the particle had an unknown and random polarisation, so that we know that we would measure either $\ket{\uparrow}$ or $\ket{\downarrow}$ upon passing it through a vertically-polarised measuring device, but we do not have any other information about its state at $t_{1}$. These are physically different situations and we are not at liberty to choose between them. \\ \\ 
There are two final differences I want to mention. First, in CH, we are free to assign a quantum state to ourselves-- indeed Gell-Mann and Hartle have used CH to explore Quantum Cosmology by assigning a quantum state to the whole universe. In RQM, we cannot assign a quantum state to ourselves, and therefore there is always implicitly a extra observer, i.e. we ourselves, assigning the quantum histories under consideration. \\
Secondly, Griffiths does not believe that quantum mechanics is contextual \cite{Griffiths2019}. However, his arguments for this are directed against a particular formulation of contextuality used by Bell. The way I am understanding contextuality in this paper is precisely to say that we can only consider compatible families, and therefore possible measurements are constrained by our choices of other measurements. Griffiths agrees with this \cite{Griffiths2017}, but believes it is a separate property which should be called, `Multitextuality'. 
\section{Discussion}
In the final part of this paper I will briefly look at two issues related to Stable Facts. The first is the claimed analogy between Special Relativity and RQM. Here, the Consistent Histories ideas can be used to clarify the similarities and differences in the way that these two theories are relational. \\
Second, the recent suggestion that RQM is augmented with, `Cross-Perspectival Links' has strong implications for the ontological meaning of Stable Facts. I therefore want to comment on this proposal. 
\subsection{Special Relativity and Quantum Mechanics}
First, I will examine how the analysis given here applies to the debates around the different ways in which Special Relativity and RQM are relational. 
An argument in favour of RQM  \cite{Rovelli1996} is the alleged similarity between the relational, observer dependent description of physical systems in that theory with the frame dependent description of physical systems in Special Relativity. For example, the way that events which are spacelike separated from two observers can be given different temporal orderings by each observer. \\
Pienaar objects \cite{Pienaar2021b} SR has a specific covariant structure, given by the underlying Minkowski Geometry; in particular the frame invariant spacetime interval between pairs of events. Relational Quantum Mechanics has no such invariant structure. Rovelli and Di Biago respond that it is precisely this which shows the true radicality of the RQM position.\\
The consistent histories analysis above can shed light on this. In Quantum Mechanics, the frameworks corresponding to different observers are in general incompatible. In SR (as for any Classical theory), this is not the case. Looking at a system from one reference frame rather than another does not alter which events happen-- hence the invariance of the underlying Minkowski spacetime, which we are simply approaching from different points of view. In QM, on the other hand, if there were  an underlying invariant structure which different measurements simply reveal from different perspectives we would be in some Hidden Variables formulation of the theory. \footnote{ The equivalent of the invariant spacetime structure in QM is, I think, twofold. First, there is the consistency of measurements (\textbf{R6} from the introduction). Another way of putting this is that the Hilbert Space structure and the rules for determining consistent and inconsistent histories remain the same for all observers. The second invariant structure in QM is the underlying laws determining the interactions.  Measurements of the same type (e.g. position, momentum, energy etc.) create the same types of states (eigenstates of the relevent operators), and all states evolve according to the Schroedinger equation when they are not being disturbed by measurement. We could think of this as a meta-formal cause in the sense that, unlike a formal cause (e.g. the shape of an enzyme which allows it to bind to certain protiens), the laws of physics do not constitute the shape of material things, but they do set the constraints for the possible Forms physical things can have. (I developed this idea by analogy with Rahner's notion of a Quasi-Formal cause \cite{Rahner1965}, though I do not think the two notions are identical).}\\
All that said, there is a clear analogy with the single framework rule in Consistent Histories when analysing paradoxes in Special Relativity. Take, for example, the twin paradox. 
In this paradox, there are two twins, one of whom remains on earth whilst the other one heads at some large fraction of the speed of light to the nearest star,  and then returns to earth.  The paradox comes about due to the fact that, since time is slower in a moving frame, the first twin (who remains on earth) should assume less time has passed for his twin (who is moving very fast relative to earth). However, the twin on the spaceship, who is at rest in his own reference frame, should also assume that less time has passed for the earthbound twin, since the earthbound twin is moving very fast relative to the spaceship, just in the opposite direction. Therefore the spaceship twin should also conclude that less time has passed for the earthbound twin. When they meet again upon the spaceship's return to earth, which of the twins will actually be younger? \\
A careful analysis of this situation is given in \cite{Lucas1990}, where the conclusion is that the paradox since the twin on the spaceship is using two reference frames-- the one on the outward journey and the one on the return journey. The issue comes from combining two inconsistent reference frames into a single description. To quote from \cite{Lucas1990}
\begin{quote}
	(The spaceship twin) changes his time reckoning in mid course... by changing the rules for dating events on earth, and so naturally gets his calculations awry. The earth-bound twin has an uninterrupted view of what is happening to his travelling brother, and so his view of the matter is undistorted 
\end{quote}
That is, the twin on earth only uses a single reference frame, and therefore has the correct account of the situation. \\ 
We can also consider the well known example of the pole moving a relativistic speeds through a barn shorter than it \cite{Taylor1966}. Suppose that the pole is 20m long in its rest frame, and the barn is 10m long. Now suppose that the pole is moving sufficiently fast relative to the barn that its length $l'$ in the barn's rest frame is 10m. Then, in the rest frame of the barn, the pole must at some point in time fit entirely in the barn; whereas in the rest frame of the pole then this cannot be the case. The paradox comes from the supposition that this is inconsistent with the fact that what is true in one reference frame must be true in all reference frames. \\
The resolution of this paradox \cite{Taylor1966} is that, since simultaneity is relative, whilst the front end of the pole exiting the barn and the rear end of the pole entering it happen at the same time in the rest frame of the barn, they do not happen at the same time in the rest frame of the pole. The front end of the pole leaves the barn in this frame before the rear end enters it. Each reference frame gives a consistent account of events-- the paradox comes from directly comparing the perspectives of observers in different Lorentz frames.\\ 
In both these cases have an analogue of the single framework rule-- we cannot directly compare events between different reference frames. However, in Special Relativity, if something is true in one reference frame, it must be true in all reference frames, whereas in Quantum Mechanics, different frameworks can correspond to different physical situations. \\
This suggests the following classification
\begin{itemize}
	\item I. Classical theories, in which the contextual aspect can be ignored-- we do not need to consider which frameworks we are using. 
	\item II. Relational theories, which posses multiple frameworks describing the same physical reality. Here we must be careful only to work in a single framework, but provided we do this then we can, `translate' from one framework to another to obtain the same physical events. As discussed above, Special Relativity is the paradigmatic example of such a theory. 
	\item III. Quantum theories, which are fully contextual in the sense that different frameworks describe possibly incompatible physical events. 
\end{itemize}
\subsection{Cross-Perspectival Links Revisited}
Recall the axiom  \textsl{Cross-perspectival links}, suggested as an addition to the RQM interpretation in \cite{Adlam2022}:
\begin{quote}
In a scenario where some observer Alice measures a variable $V$ of a system $S$, then provided Alice does not undergo any interactions which destroy the information about $V$ stored in Alice's physical variables, if Bob subsequently measures the physical variable representing Alice's information about the variable $V$, then Bob's measurement result will match Alice's measurement result. 
\end{quote}
In this section I will suggest an alternative way to derive the axiom, and discuss the differences between this approach and the original one \cite{Adlam2022}. We start with the 1996 postulates about RQM in terms of information: \cite{Rovelli1996}
\begin{quote}
	\begin{enumerate}
		\item There is a maximum amount of relevant information which can be extracted from a system
 \item It is always possible to acquire new information about a system. 
 	\end{enumerate}
\end{quote}    
I suggest that instead of CPL, we can add the axiom:
\begin{quote}
	\textbf{Creative Measurement\  (CM)}: Once information about a system is created in an interaction, the system possess that information until the next interaction it is involved in, which may preserve, alter or destroy that information
	\footnote{There is an interesting parallel here to Newton's first law, which is worth exploring.}
\end{quote}
When combined with the 1996 axioms, this implies that once the values of a variable is created through an interaction\footnote{The idea that measurement is creative and changes the things measured goes right back to Heisenberg \cite{SEPUncert}. A good example is  given by Feynman's discussion of the double slit experiment \cite{FeynmannLect}. Feynman points out that we can only detect which slit the particle goes through by closing one or the other of them (as the detector blocks the particle). When we only partially close the slit (and only sometimes detect the particle), the interference pattern re-emerges in inverse proportion to how closed the slit is, and hence how much we obstruct the particle.}  , if the next interaction is compatible, then the information about the value, and hence the value itself, remains unchanged. If the next interaction is incompatible, then the information about the value of the variable is lost and the value-- or even the variable-- is changed\footnote{ for example, a system in state $\ket{\uparrow}$, when measured in a horizontal polarizer, loses the vertical polarization variable and instead gains a horizontal state, either $\ket{\leftarrow}$ or $\ket{\rightarrow}$}. \\
The main difference between this view and the one in \cite{Adlam2022} is the underlying ontology. Adlam and Rovelli explicitly deny that variables have values in between interactions, saying, 
\begin{quote}
	\textsl{Systems do note have observer-independent ontic states which persist through time storing their variables:\ variables are only ever defined instantaneously}
\end{quote}
They are also clear that 
\begin{quote}
	\textsl{The version of RQM we have presented here exhibits gappy metaphysical indeterminacy, since variables have no definite values in regions between quantum events}
\end{quote}
This standard RQM approach is often called a ,`Sparse Flash' ontology, where the variables only take values instantaneously. In contrast, what I am proposing could be defined as a, `beam ontology', where variables and their values are created at an interaction, and persist until the next interaction.\\
Note this approach is still relational-- there are no properties independent of interactions between systems. \footnote{Having set out this way of understanding RQM, I want to be clear as to why this is not simply a hidden variables theory. Hidden Variables are variables in a quantum theory which always have  a definite value,  which is revealed by the measurement of that variable.  A key feature of hidden variables accounts of Quantum Mechanics is that the measurement of a variable is the same as the value before the measurement-- this  is called \textsl{Faithful Measurement}\cite{SEPKS} and this is certainly not satisfied in the account I am giving here. This is an example of the noncontextuality of RQM, discussed in the introduction. It should be noted here that the only Hidden Variables theory which agrees with the predictions of Quantum Mechanics is Bohmian mechanics, in which only the positions and momenta are hidden variables in this way}How do we get from this to CPL? Well, if Alice has measured a variable $V$ of $S$, then by CM,  $S$ continues to have that variable until it undergoes an interaction which destroys that information. In addition, Alice will have measured $V$, and this measurement will remain unless Alice has undergone some interaction destroying her measurement result.  If Bob believes that neither S nor Alice have  undergone such an interaction, then he is justified in assuming that his measurement of $V$ will match Alice's -- which is precisely the content of CPL. \\
Whilst these two ontologies are different, they will both give the mathematical structure of RQM. One advantage, as I see it, of the, `beam' ontology is that it avoids the metaphysical indeterminacy of the, `sparse flash' ontology. Another is that is seems truer to the founding principle of RQM, that, `information is physical'\cite{Rovelli1996}-- rather than being an axiom about observing systems which introduces a non- relational element into RQM, it is an axiom about information, which then gives the CPL axiom as a consequence.  That said,  more work will be needed to examine the pros and cons of the two positions. Both ontological views strongly benefit from the Consistent Histories methods I have presented in this paper, as these give a method for determining precisely which interactions satisfy the condition, `provided Alice does not undergo any interactions which destroy the information about $V$', and hence when CPL applies. 
\section{Conclusion}
In this paper I have looked at Stable Facts in RQM, and attempted to give a rigorous method for determining which facts are stable, by using the mathematical formalism of the Consistent Histories method.\\ 
The heart of this formalism is Quantum Reasoning, which is a  Quantum Logic based on subspaces and operators on a Hilbert Space, as opposed to the usual classical logic based on subsets of a classical phase space. This allows us to view the Hilbert Space structure of Quantum Mechanics as a non-standard logical framework for analysing situations in which the truth of propositions depends upon the ordering of these propositions. This logic allows propositions not only to be true or false, but to be undefined\footnote{Such logics are known in the philosophy of logic as, `paraconsistent'; and have antecedents as far back as Plato's \textsl{Sophist}\cite{Haecker2022}}, and propositions are defined or undefined relative to the other propositions they are considered with. We can therefore see this as a contextual logic, capturing the contextual nature of Quantum Mechanics. In fact, Quantum Mechanics is simply this contextual logic applied to physics \footnote{It would be interesting to try and apply this logic to model other areas where contextuality might apply}. We can see classical theories as special cases of quantum theories in which all the operators commute. \footnote{This implements the insight of Rovelli \cite{Rovelli1996} that all interactions can be described by Quantum Theory. }.\\ 
Why does this give RQM? Because according to this analysis, systems have properties only relative to a particular history of interactions, with respect to a particular system.  Some measurements preserve the information which has already been gained about a system about an observer, and others destroy it. Likewise, some information can be shared between observers, if their measurements are compatible; and other information cannot be shared,if their measurements are mutually destructive. \\
This is where methods in this paper are useful, by giving a rigorous process for determining when facts are stable between system, and when they are relative. This could also lead to a renewed focus on the information-based approach to RQM, which has been neglected in recent years. There are many opportunities for further work in developing the application of consistent histories to RQM, and in thinking through the consequences for the ontological foundations of the interpretation. 
\bibliography{Jan2023.bib}


\begin{thebibliography}{38}
\ifx \bisbn   \undefined \def \bisbn  #1{ISBN #1}\fi
\ifx \binits  \undefined \def \binits#1{#1}\fi
\ifx \bauthor  \undefined \def \bauthor#1{#1}\fi
\ifx \batitle  \undefined \def \batitle#1{#1}\fi
\ifx \bjtitle  \undefined \def \bjtitle#1{#1}\fi
\ifx \bvolume  \undefined \def \bvolume#1{\textbf{#1}}\fi
\ifx \byear  \undefined \def \byear#1{#1}\fi
\ifx \bissue  \undefined \def \bissue#1{#1}\fi
\ifx \bfpage  \undefined \def \bfpage#1{#1}\fi
\ifx \blpage  \undefined \def \blpage #1{#1}\fi
\ifx \burl  \undefined \def \burl#1{\textsf{#1}}\fi
\ifx \doiurl  \undefined \def \doiurl#1{\url{https://doi.org/#1}}\fi
\ifx \betal  \undefined \def \betal{\textit{et al.}}\fi
\ifx \binstitute  \undefined \def \binstitute#1{#1}\fi
\ifx \binstitutionaled  \undefined \def \binstitutionaled#1{#1}\fi
\ifx \bctitle  \undefined \def \bctitle#1{#1}\fi
\ifx \beditor  \undefined \def \beditor#1{#1}\fi
\ifx \bpublisher  \undefined \def \bpublisher#1{#1}\fi
\ifx \bbtitle  \undefined \def \bbtitle#1{#1}\fi
\ifx \bedition  \undefined \def \bedition#1{#1}\fi
\ifx \bseriesno  \undefined \def \bseriesno#1{#1}\fi
\ifx \blocation  \undefined \def \blocation#1{#1}\fi
\ifx \bsertitle  \undefined \def \bsertitle#1{#1}\fi
\ifx \bsnm \undefined \def \bsnm#1{#1}\fi
\ifx \bsuffix \undefined \def \bsuffix#1{#1}\fi
\ifx \bparticle \undefined \def \bparticle#1{#1}\fi
\ifx \barticle \undefined \def \barticle#1{#1}\fi
\bibcommenthead
\ifx \bconfdate \undefined \def \bconfdate #1{#1}\fi
\ifx \botherref \undefined \def \botherref #1{#1}\fi
\ifx \url \undefined \def \url#1{\textsf{#1}}\fi
\ifx \bchapter \undefined \def \bchapter#1{#1}\fi
\ifx \bbook \undefined \def \bbook#1{#1}\fi
\ifx \bcomment \undefined \def \bcomment#1{#1}\fi
\ifx \oauthor \undefined \def \oauthor#1{#1}\fi
\ifx \citeauthoryear \undefined \def \citeauthoryear#1{#1}\fi
\ifx \endbibitem  \undefined \def \endbibitem {}\fi
\ifx \bconflocation  \undefined \def \bconflocation#1{#1}\fi
\ifx \arxivurl  \undefined \def \arxivurl#1{\textsf{#1}}\fi
\csname PreBibitemsHook\endcsname

\bibitem[\protect\citeauthoryear{Rovelli}{1996}]{Rovelli1996}
\begin{barticle}
\bauthor{\bsnm{Rovelli}, \binits{C.}}:
\batitle{Relational quantum mechanics}.
\bjtitle{Int. J. Theo. Phys.}
\bvolume{35},
\bfpage{1637}--\blpage{1678}
(\byear{1996})
\end{barticle}
\endbibitem

\bibitem[\protect\citeauthoryear{{van Fraassen}}{2010}]{VanFraassen2010}
\begin{barticle}
\bauthor{\bsnm{{van Fraassen}}, \binits{B.C.}}:
\batitle{Rovelli's world}.
\bjtitle{Foundations of Physics}
\bvolume{40},
\bfpage{390}--\blpage{417}
(\byear{2010})
\end{barticle}
\endbibitem

\bibitem[\protect\citeauthoryear{Laudisa}{2019}]{Laudisa2019}
\begin{barticle}
\bauthor{\bsnm{Laudisa}, \binits{F.}}:
\batitle{Open problems in relational quantum mechanics}.
\bjtitle{Journal for General Philosophy of Science}
\bvolume{50},
\bfpage{215}--\blpage{230}
(\byear{2019})
\end{barticle}
\endbibitem

\bibitem[\protect\citeauthoryear{Di~Biago and Rovelli}{2021}]{DiBiagio2021}
\begin{botherref}
\oauthor{\bsnm{Di~Biago}, \binits{A.}},
\oauthor{\bsnm{Rovelli}, \binits{C.}}:
Stable facts, relative facts.
Foundations of Physics
\textbf{51}(30)
(2021)
\end{botherref}
\endbibitem

\bibitem[\protect\citeauthoryear{Di~Biagio and Rovelli}{2022}]{DiBiagio2022}
\begin{botherref}
\oauthor{\bsnm{Di~Biagio}, \binits{A.}},
\oauthor{\bsnm{Rovelli}, \binits{C.}}:
Quantum mechanics is about facts not states: A reply to pienaar and brukner.
Foundations of Physics
(62)
(2022)
\end{botherref}
\endbibitem

\bibitem[\protect\citeauthoryear{Brukner}{2021}]{Brukner2021}
\begin{botherref}
\oauthor{\bsnm{Brukner}, \binits{C.}}:
Qubits are not observers-- a no-go theorem.
preprint
(2021)
\end{botherref}
\endbibitem

\bibitem[\protect\citeauthoryear{Pienaar}{2021}]{Pienaar2021a}
\begin{botherref}
\oauthor{\bsnm{Pienaar}, \binits{J.}}:
Qbism and relational quantum mechanics compared.
Foundations of Physics
\textbf{51}
(2021)
\end{botherref}
\endbibitem

\bibitem[\protect\citeauthoryear{J.Pienaar}{2021}]{Pienaar2021b}
\begin{botherref}
\oauthor{\bsnm{J.Pienaar}}:
A quintet of quandries: five no-go theorems for relational quantum mechanics.
Foundations of Physics
\textbf{51}
(2021)
\end{botherref}
\endbibitem

\bibitem[\protect\citeauthoryear{Lawrence et~al.}{2023}]{Lawrence2023}
\begin{botherref}
\oauthor{\bsnm{Lawrence}, \binits{J.}},
\oauthor{\bsnm{Markiewicz}, \binits{M.}},
\oauthor{\bsnm{Zukowski}, \binits{M.}}:
Relative facts of relational quantum mechanics are incompatiable with quantum
  mechanics.
Quantum
\textbf{7}
(2023)
\end{botherref}
\endbibitem

\bibitem[\protect\citeauthoryear{Cavalcanti et~al.}{}]{Cavalcanti2023}
\begin{botherref}
\oauthor{\bsnm{Cavalcanti}, \binits{E.G.}},
\oauthor{\bsnm{DiBiagio}, \binits{A.}},
\oauthor{\bsnm{Rovelli}, \binits{C.}}:
On the consistency of relative facts.
https://arxiv.org/abs/2305.07343
\end{botherref}
\endbibitem

\bibitem[\protect\citeauthoryear{Muci{\~n}o et~al.}{2022}]{Mucino2022}
\begin{botherref}
\oauthor{\bsnm{Muci{\~n}o}, \binits{R.}},
\oauthor{\bsnm{Okon}, \binits{E.}},
\oauthor{\bsnm{Sudarsky}, \binits{D.}}:
Assessing relational quantum mechanics.
Synthese
\textbf{200}(399)
(2022)
\end{botherref}
\endbibitem

\bibitem[\protect\citeauthoryear{Smerlak and Rovelli}{2007}]{Smerlak2007}
\begin{barticle}
\bauthor{\bsnm{Smerlak}, \binits{M.}},
\bauthor{\bsnm{Rovelli}, \binits{C.}}:
\batitle{Relational {EPR}}.
\bjtitle{Foundations of Physics}
\bvolume{37},
\bfpage{427}--\blpage{445}
(\byear{2007})
\end{barticle}
\endbibitem

\bibitem[\protect\citeauthoryear{Martin-Dussaud et~al.}{2019}]{Dussaud2019}
\begin{barticle}
\bauthor{\bsnm{Martin-Dussaud}, \binits{P.}},
\bauthor{\bsnm{Rovelli}, \binits{C.}},
\bauthor{\bsnm{Zalamea}, \binits{F.}}:
\batitle{The notion of locality in relational quantum mechanics}.
\bjtitle{Foundations of Physics}
\bvolume{49},
\bfpage{96}--\blpage{106}
(\byear{2019})
\end{barticle}
\endbibitem

\bibitem[\protect\citeauthoryear{C.Rovelli}{2022}]{Rovelli2022}
\begin{bchapter}
\bauthor{\bsnm{C.Rovelli}}:
\bctitle{The relational interpretation of quantum mechanics}.
In: \bbtitle{The Oxford Handbook of the History of Interpretation of Quantum
  Physics (preprint on ArXiv)}.
\bpublisher{OUP},
\blocation{{}}
(\byear{2022})
\end{bchapter}
\endbibitem

\bibitem[\protect\citeauthoryear{Griffiths}{2008}]{Griffiths2008}
\begin{bbook}
\bauthor{\bsnm{Griffiths}, \binits{R.B.}}:
\bbtitle{Consistent Quantum Mechanics}.
\bpublisher{CUP},
\blocation{{}}
(\byear{2008})
\end{bbook}
\endbibitem

\bibitem[\protect\citeauthoryear{Griffiths}{2014}]{Griffiths2014}
\begin{botherref}
\oauthor{\bsnm{Griffiths}, \binits{R.B.}}:
The new quantum logic.
Foundations of Physics
\textbf{44}
(2014)
\end{botherref}
\endbibitem

\bibitem[\protect\citeauthoryear{Adlam and Rovelli}{2023}]{Adlam2022}
\begin{botherref}
\oauthor{\bsnm{Adlam}, \binits{E.}},
\oauthor{\bsnm{Rovelli}, \binits{C.}}:
Information is physical: Cross-perspectival links in relational quantum
  mechanics.
Philosophy of Physics
\textbf{1}(1)
(2023)
\end{botherref}
\endbibitem

\bibitem[\protect\citeauthoryear{Frauchiger and Renner}{2018}]{Frauchiger2018}
\begin{botherref}
\oauthor{\bsnm{Frauchiger}, \binits{D.}},
\oauthor{\bsnm{Renner}, \binits{R.}}:
Quantum theory cannot consistently describe the use of itself.
Nat. Commun.
\textbf{9}(1)
(2018)
\end{botherref}
\endbibitem

\bibitem[\protect\citeauthoryear{C.Held}{2000, rev. 2022}]{SEPKS}
\begin{bchapter}
\bauthor{\bsnm{C.Held}}:
\bctitle{The {Kochen}-{Specker} theorem}.
In: \bbtitle{SEP}.
\bpublisher{Stanford},
\blocation{{}}
(\byear{2000, rev. 2022})
\end{bchapter}
\endbibitem

\bibitem[\protect\citeauthoryear{Dirac}{1988}]{Dirac1988}
\begin{bbook}
\bauthor{\bsnm{Dirac}, \binits{P.A.M.}}:
\bbtitle{Principles of Quantum Mechanics},
\bedition{4th} edn.
\bpublisher{OUP},
\blocation{{}}
(\byear{1988})
\end{bbook}
\endbibitem

\bibitem[\protect\citeauthoryear{Feynman}{2010}]{FeynmannLect}
\begin{bbook}
\bauthor{\bsnm{Feynman}, \binits{R.}}:
\bbtitle{Feynman Lectures on Physics},
\bedition{New millennium edition} edn.
\bpublisher{Basic Books},
\blocation{{}}
(\byear{2010})
\end{bbook}
\endbibitem

\bibitem[\protect\citeauthoryear{E.P.Wigner}{1960}]{Wigner1960}
\begin{botherref}
\oauthor{\bsnm{E.P.Wigner}}:
The unreasonable effectiveness of mathematics in the natural sciences.
Commun. Pure. Appl. Math.
\textbf{13}(1)
(1960)
\end{botherref}
\endbibitem

\bibitem[\protect\citeauthoryear{Oldofredi and Calosi}{2021}]{Oldofredi2021}
\begin{botherref}
\oauthor{\bsnm{Oldofredi}, \binits{A.}},
\oauthor{\bsnm{Calosi}, \binits{C.}}:
Relational quantum mechanics and the {PBR} theorem: A peaceful coexistence.
Foundations of Physics
\textbf{51}(82)
(2021)
\end{botherref}
\endbibitem

\bibitem[\protect\citeauthoryear{R.Penrose}{1971}]{Penrose1971}
\begin{bchapter}
\bauthor{\bsnm{R.Penrose}}:
\bctitle{Angular momentum: An approach to combinatorial space-time}.
In: \bbtitle{Quantum Theory and Beyond},
pp. \bfpage{151}--\blpage{180}.
\bpublisher{CUP},
\blocation{{}}
(\byear{1971})
\end{bchapter}
\endbibitem

\bibitem[\protect\citeauthoryear{Griffiths}{2014, rev. 2019}]{GriffithsSEP}
\begin{bchapter}
\bauthor{\bsnm{Griffiths}, \binits{R.B.}}:
\bctitle{The consistent histories approach to quantum mechanics}.
In: \bbtitle{SEP}.
\bpublisher{Stanford},
\blocation{{}}
(\byear{2014, rev. 2019})
\end{bchapter}
\endbibitem

\bibitem[\protect\citeauthoryear{Griffiths}{1984}]{Griffiths1984}
\begin{barticle}
\bauthor{\bsnm{Griffiths}, \binits{R.B.}}:
\batitle{Consistent histories and the interpretation of quantum mechanics}.
\bjtitle{Journal of Statistical Physics}
\bvolume{36},
\bfpage{219}--\blpage{272}
(\byear{1984})
\end{barticle}
\endbibitem

\bibitem[\protect\citeauthoryear{Omn{\'e}s}{1987}]{Omnes1987}
\begin{barticle}
\bauthor{\bsnm{Omn{\'e}s}, \binits{R.}}:
\batitle{Interpretation of quantum mechanics}.
\bjtitle{Physics Letters A}
\bvolume{125}(\bissue{4}),
\bfpage{169}--\blpage{172}
(\byear{1987})
\end{barticle}
\endbibitem

\bibitem[\protect\citeauthoryear{Gell-Mann and Hartle}{1990}]{GellMan1990}
\begin{bchapter}
\bauthor{\bsnm{Gell-Mann}, \binits{M.}},
\bauthor{\bsnm{Hartle}, \binits{J.B.}}:
\bctitle{Quantum mechanics in light of quantum cosmology}.
In: \bbtitle{Complexity, Entropy and the Physics of Information},
pp. \bfpage{321}--\blpage{343}.
\bpublisher{Addison Wesley},
\blocation{{}}
(\byear{1990})
\end{bchapter}
\endbibitem

\bibitem[\protect\citeauthoryear{Rocha et~al.}{2022}]{Rocha2022}
\begin{bchapter}
\bauthor{\bsnm{Rocha}, \binits{G.R.}},
\bauthor{\bsnm{Rickles}, \binits{D.}},
\bauthor{\bsnm{Boge}, \binits{F.J.}}:
\bctitle{A brief historical perspective on the consistent histories
  interpretation of quantum mechanics}.
In: \beditor{\bsnm{Freire~Jr.}, \binits{O.}} (ed.)
\bbtitle{The Oxford Handbook of the History of Interpretation of Quantum
  Physics (preprint on ArXiv)}.
\bpublisher{OUP}, \blocation{???}
(\byear{2022})
\end{bchapter}
\endbibitem

\bibitem[\protect\citeauthoryear{Griffiths}{1996}]{Griffiths1996}
\begin{botherref}
\oauthor{\bsnm{Griffiths}, \binits{R.B.}}:
Consistent histories and quantum reasoning.
Phys. Rev. A.
\textbf{54}(4)
(1996)
\end{botherref}
\endbibitem

\bibitem[\protect\citeauthoryear{Birkhoff and {von
  Neumann}}{1936}]{Neumann1936}
\begin{barticle}
\bauthor{\bsnm{Birkhoff}, \binits{G.}},
\bauthor{\bsnm{{von Neumann}}, \binits{J.}}:
\batitle{The logic of quantum mechanics}.
\bjtitle{Annals of Mathematics}
\bvolume{37},
\bfpage{823}--\blpage{843}
(\byear{1936})
\end{barticle}
\endbibitem

\bibitem[\protect\citeauthoryear{Griffiths}{2017}]{Griffiths2017}
\begin{botherref}
\oauthor{\bsnm{Griffiths}, \binits{R.B.}}:
What quantum measurements measure.
Phys. Rev. A.
\textbf{96}
(2017)
\end{botherref}
\endbibitem

\bibitem[\protect\citeauthoryear{Griffiths}{2019}]{Griffiths2019}
\begin{botherref}
\oauthor{\bsnm{Griffiths}, \binits{R.B.}}:
Quantum measurements and contextuality.
Phil. Trans. Roy. Soc. A.
\textbf{377}
(2019)
\end{botherref}
\endbibitem

\bibitem[\protect\citeauthoryear{Rahner}{1965}]{Rahner1965}
\begin{bchapter}
\bauthor{\bsnm{Rahner}, \binits{K.}}:
\bctitle{Nature and {G}race}.
In: \bbtitle{Theological Investigations},
\bedition{2nd} edn.
\bpublisher{Darton, Longmann and Todd Ltd},
\blocation{{}}
(\byear{1965})
\end{bchapter}
\endbibitem

\bibitem[\protect\citeauthoryear{Lucas and Hodgeson}{1990}]{Lucas1990}
\begin{bbook}
\bauthor{\bsnm{Lucas}, \binits{J.R.}},
\bauthor{\bsnm{Hodgeson}, \binits{P.E.}}:
\bbtitle{Spacetime and Electromagnetism}.
\bpublisher{OUP},
\blocation{{}}
(\byear{1990})
\end{bbook}
\endbibitem

\bibitem[\protect\citeauthoryear{Taylor and Wheeler}{1966}]{Taylor1966}
\begin{bbook}
\bauthor{\bsnm{Taylor}, \binits{E.F.}},
\bauthor{\bsnm{Wheeler}, \binits{J.A.}}:
\bbtitle{Spacetime Physics},
\bedition{2nd} edn.
\bpublisher{W.H.Freeman and Co.},
\blocation{{}}
(\byear{1966})
\end{bbook}
\endbibitem

\bibitem[\protect\citeauthoryear{Hilgevoord}{2001, Rev. 2016}]{SEPUncert}
\begin{bchapter}
\bauthor{\bsnm{Hilgevoord}, \binits{J.}}:
\bctitle{The uncertainty principle}.
In: \bbtitle{SEP}.
\bpublisher{Stanford},
\blocation{{}}
(\byear{2001, Rev. 2016})
\end{bchapter}
\endbibitem

\bibitem[\protect\citeauthoryear{Haecker}{2022}]{Haecker2022}
\begin{botherref}
\oauthor{\bsnm{Haecker}, \binits{R.}}:
Splitting the difference: Contradiction and the {Trinity} in {Plato's}
  `{Sophist}'.
Macrina
(2022)
\end{botherref}
\endbibitem

\end{thebibliography}

\end{document}